\documentclass[epj]{webofc}
\usepackage[utf8]{inputenc}
\usepackage[varg]{txfonts}   
\usepackage{booktabs}
\usepackage{bm}
\usepackage{xcolor}
\definecolor{darkred}{rgb}{0.4,0.0,0.0}
\definecolor{darkgreen}{rgb}{0.0,0.4,0.0}
\definecolor{darkblue}{rgb}{0.0,0.0,0.4}
\usepackage[bookmarks,linktocpage,colorlinks,
    linkcolor = darkred,
    urlcolor  = darkblue,
    citecolor = darkgreen]{hyperref}
\usepackage{subfigure}

\wocname{EPJ Web of Conferences}
\woctitle{Lattice2017}

\begin{document}

\selectlanguage{english}

\title{%
An estimate for the thermal photon rate from lattice QCD
}

\author{%
\firstname{Bastian B.}  \lastname{Brandt}\inst{1}\and
\firstname{Anthony}  \lastname{Francis}\inst{2}\and
\firstname{Tim} \lastname{Harris}\inst{3}\fnsep\thanks{Speaker, \email{harris@him.uni-mainz.de}}\and
\firstname{Harvey B.}  \lastname{Meyer}\inst{3,4}\and\newline
\firstname{Aman} \lastname{Steinberg}\inst{4}\fnsep\thanks{Speaker, \email{amsteinb@uni-mainz.de}}
}

\institute{%
Institute for Theoretical Physics, Goethe University, 60438 Frankfurt am Main, Germany
\and
Dept. of Physics and Astronomy, York University, Toronto, Ontario, M3J 1P3, Canada
\and
Helmholtz-Institut Mainz, Johannes Gutenberg-Universit\"at, 55099 Mainz, Germany
\and
PRISMA Cluster of Excellence and Institute of Nuclear Physics,\\
Johannes Gutenberg-Universit\"at, 55128 Mainz, Germany
}

\abstract{%
We estimate the production rate of photons by the quark-gluon plasma
in lattice QCD.
We propose a new correlation function which provides better control over
the systematic uncertainty in estimating the photon production rate at photon
momenta in the range ${\pi T}/{2}$ to $2\pi T$.
The relevant Euclidean vector current correlation functions are computed with
$N_\mathrm f =2$ Wilson clover fermions in the chirally-symmetric
phase.
In order to estimate the photon rate, an ill-posed problem for the
vector-channel spectral function must be regularized.
We use both a direct model for the spectral function and a model-independent
estimate from the Backus-Gilbert method to give an estimate for
the photon rate.}

\maketitle

\section{Introduction}
\label{sec:intro}
Lattice QCD has had a tremendous impact in computing the equilibrium properties of strongly interacting matter at non-zero temperature. Its real-time properties, however, are difficult to compute
from numerical lattice calculations because they require the analytic
continuation of incomplete and noisy Euclidean two-point functions.
Yet, the dynamical parameters of thermal systems, like transport coefficients
and production rates of weakly-coupled external currents, are needed to
interpret, for example, the equilibration of the thermal medium formed after a
heavy-ion collision~\cite{Shen:2013vja} or the abundance of particles in the
early universe~\cite{Asaka:2006rw}. In addition, they are interesting in their own right to characterize the medium.

In principle, all of the dynamical properties, such as quasiparticle widths and
diffusion coefficients of conserved charges, are encoded in 
temperature-dependent spectral functions which relate all real- and
imaginary-time two-point correlation functions.
The spectral representation of the Euclidean correlation function of the electromagnetic current,
$V^\mathrm{em}_\mu(x)=\sum_fQ_f\overline\psi_f(x)\gamma_\mu\psi_f(x)$ where
$\{\gamma_\mu,\gamma_\nu\}=2\eta_{\mu\nu}=2\,{\rm diag}(+,-,-,-)$, reads\footnote{Here we assume that 
the number of temporal indices among $\mu,\nu$ is not exactly one, in which case the kernel is different.}
\begin{align}
\label{eq:specrep}
    G_{\mu\nu}(\tau,\bm k) &= \int_0^\infty\frac{\mathrm d\omega}{2\pi}\,
        K(\tau,\omega)\rho_{\mu\nu}(\omega, \bm k),
        \quad\textrm{where}\quad
        K(\tau,\omega)=\frac{\cosh(\omega\tau-\omega\beta/2)}{\sinh(\beta\omega/2)}.
\end{align}
It represents an ill-posed inverse problem 
when the correlator is finitely sampled, even though definite algorithms for this task exist~\cite{Cuniberti:2001hm}.
Nevertheless, there may exist particular observables which can be constrained by
the Euclidean observables.

Here we propose a new correlator whose spectral function has
no UV growth, from which the production rate of real photons from a
thermal medium of quarks and gluons can be computed.
Furthermore, we present preliminary results of the photon rate using a 
continuum-extrapolated correlator computed with $N_\mathrm f=2$ flavours of
O($a$)-improved Wilson fermions.

\subsection{The photon production rate}
\label{sec:phorate}
The differential production rate of weakly-coupled massless vector bosons 
from a thermal medium of gluons and quarks in the flavour-symmetric limit
is given, to leading order in the electromagnetic coupling $\alpha_\mathrm{em}$,
by~\cite{McLerran:1984ay}
\begin{align}
    \label{eq:phorate}
    k\frac{\mathrm d\Gamma_\gamma}{\mathrm d^3k}&=
    \left(\sum_f Q_f^2\right)\frac{\alpha_\mathrm{em}}{4\pi^2}
    n_\mathrm{B}(\omega=k)\rho^\mu\,_\mu(\omega=\left|\bm k\right|, \bm k),
\end{align}
where $n_\mathrm B(\omega)=1/[\exp(\beta\omega)-1]$ is the Bose-Einstein distribution.
The rate is therefore determined by the spectral function evaluated on the
photon mass-shell.
This rate can be directly compared to the measurement of direct non-prompt
real photons, i.e. those not originating from hadronic decays or initial
partonic scatterings, which have been observed at
RHIC~\cite{Afanasiev:2012dg} and the LHC~\cite{delaCruz:2012ru,Milov:2012pd}.

Theoretically, this rate has been computed in QCD at
weak-coupling~\cite{Arnold:2001ms,Ghiglieri:2013gia}, and QCD-like theories at
strong-coupling~\cite{CaronHuot:2006te}.
These two scenarios show qualitatively different behaviour in the soft
limit $k\equiv\left|\bm k\right|\to0$, where the weak-coupling prediction for $\rho^{\mu}{}_\mu(k,\bm k)/k$
is of order $1/\alpha_\mathrm s^2$, while the strong-coupling prediction is 
independent of the coupling.
It would therefore be desirable to have a lattice prediction in the region of
photon momenta of a few times the temperature to distinguish these
possibilities, and further elucidate the nature of the medium itself.
Pioneering numerical calculations have been performed in pure gauge
theory~\cite{Ghiglieri:2016tvj}, which utilized advanced perturbative results on the
UV behaviour of the Euclidean correlator $G^\mu\,_\mu(\tau,\bm k)$, 
closely-related to the dilepton production rate, to constrain the photon rate.

\subsection{UV-finite correlator}
\label{sec:uvcorr}
As a consequence of the continuum vector Ward identity $\omega^2\rho_{00}=k^ik^j\rho_{ij}$, on the light cone
\begin{align}
    \label{eq:lightcone}
    k^2\rho_{00}(\omega,\bm k) - k^ik^j\rho_{ij}(\omega,\bm k)
      &= 0, \qquad \omega=\left|\bm k\right|\equiv k.
\end{align}
It is therefore helpful to consider the linear combination
\begin{align}
    \label{eq:rholambda}
    \rho_\lambda(\omega,\bm k) &\equiv \left(\delta^{ij} - \textstyle{\frac{k^ik^j}{k^2}}\right)
            \rho_{ij}(\omega,\bm k)
            - {\lambda}\left(\rho_{00}(\omega,\bm k) -
            \textstyle{\frac{k^ik^j}{k^2}}\rho_{ij}(\omega,\bm k)
           \right).
\end{align}
The spectral function with $\lambda=1$ corresponds to   $\rho_1(\omega, \bm k)=\rho^\mu\,_\mu(\omega,\bm k)$,
directly relevant to the dilepton rate.
However, this channel contains the large ultraviolet contribution of order $\omega^2$ at large $\omega$;
at small $k$, it also contains the diffusion pole. To avoid confronting these complications simultaneously, 
one can analyze separately 
\begin{enumerate}
\item the purely transverse case $\lambda=0$, which does not couple to the diffusion pole;
\item the case $\lambda=-2$, corresponding to the difference of the transverse and longitudinal channels.
\end{enumerate}
The latter linear combination vanishes identically in the vacuum and is highly suppressed in the ultraviolet.
Here we concentrate on the case $\lambda=-2$; in the future, we plan to also analyze the case $\lambda=0$,
which should yield consistent photon rates, thus providing a powerful cross-check. At the end, the spectral functions 
with $\lambda=0$ and $\lambda=-2$ can be recombined in order to predict the dilepton rate.
The importance of removing UV divergences from Euclidean correlators to
estimate thermal real-time observables has also been discussed in ref.~\cite{Burnier:2012ts}.

\begin{figure}[h]
    \centering
    \includegraphics[scale=0.7]{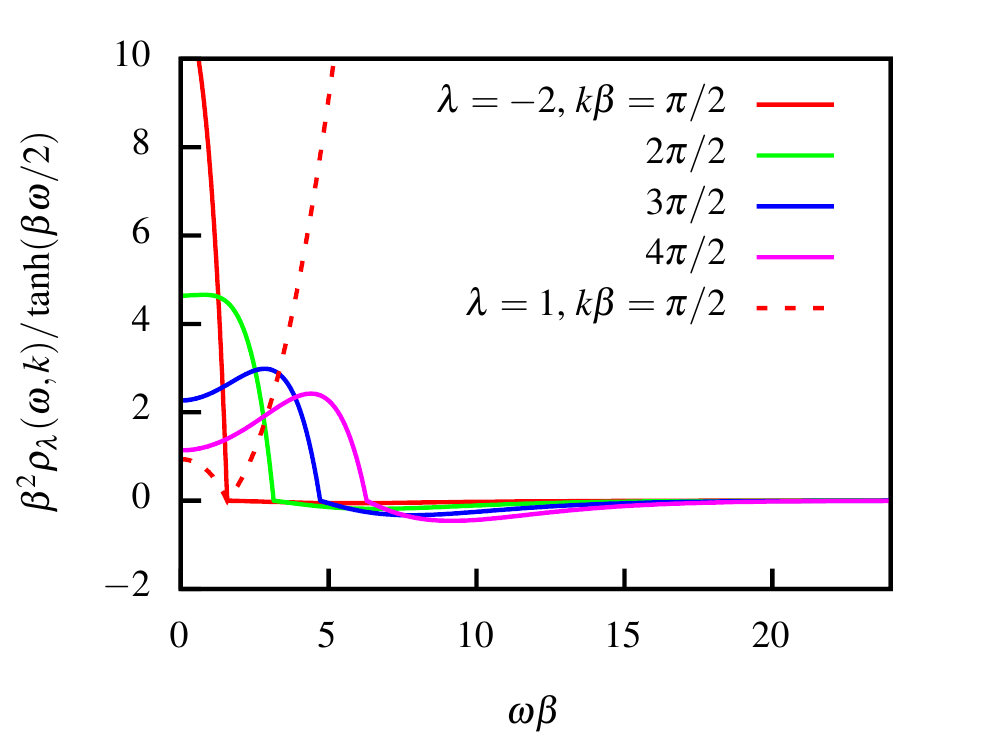}
    \caption{The spectral function $\rho_\lambda(\omega,\bm k)$ computed in
        tree-level continuum perturbation theory, illustrating the improved UV
        behaviour of the $\lambda=-2$ spectral function (solid lines) versus
        the standard divergent choice $\lambda=1$ (dashed line).}
    \label{fig:rho_lo}
\end{figure}

Figure~\ref{fig:rho_lo} illustrates the effect of the cancellation on
the tree-level spectral function in the solid lines.  The standard
correlation function ($\lambda=1$) is shown for the lowest momenta in
the dashed line, which diverges as $\omega^2$ at large frequencies.
The spectral function with $\lambda=-2$ on the other hand is very
suppressed for $\omega>k$, thus making this channel very sensitive to
the infrared physics of interest.  Note that the spectral function
evaluated on the photon mass-shell (at the kink), and thus the photon rate,
vanishes at this order in perturbation theory. 
If one thinks of the inverse problem as resulting in a `smearing' of the
actual spectral function, as is
explicitly the case in the Backus-Gilbert method, then this represents
a difficulty, since the spectral weight is of order unity for
$\omega\lesssim k$.

\section{Continuum limit}
\label{sec:contlimit}
We have generated a series of ensembles to take the continuum limit at a
single temperature, approximately $T=250\,\textrm{MeV}$, above the crossover to the chirally symmetric phase, and
an additional ensemble at a single lattice spacing deep in the deconfined
phase, approximately $T=500\,\mathrm{MeV}$; see table~\ref{tab:ensembles}.
We use the non-perturbatively O($a$)-improved Wilson action~\cite{Jansen:1997nc} with
$N_\mathrm f=2$ Wilson fermions and the Wilson gauge action.
The parameters were chosen using the running of the coupling and quark masses
as determined by the CLS collaboration~\cite{Fritzsch:2012wq}. The lattice
with $N_\tau\equiv\beta/a=16$ at $T\approx 250\,\mathrm{MeV}$, where
$\beta=T^{-1}$ is the inverse temperature, has been used for our previous studies,
see refs.~\cite{Brandt:2012jc,Brandt:2013faa,Brandt:2015aqk}.
\begin{table}
    \centering
    \resizebox{0.6\textwidth}{!}{%
    \begin{tabular}{ccccccc}
        \toprule
        $T$ (MeV)   &   $T/T_\mathrm{c}$  &   $\beta_\mathrm{LAT}$    &
        $\beta/a$   &   $L/a$    &   $m_{\mathrm{\overline{MS}(2\,GeV)}}$ (MeV) & $N_\mathrm{meas}$\\
        \midrule
        250 &   1.2 &   5.3     &   12      &     48 & 12 & 8256\\
        ''  &   ''  &   5.5     &   16      &     64 & '' & 4880\\
        ''  &   ''  &   5.83    &   24      &     96 & '' & 1680\\
        \midrule
        500 & 2.4 &   6.04    &   16      &     64 & '' & 8064\\
        \bottomrule
    \end{tabular}}
    \caption{The details of the ensembles used in this work.
             The tuning of the lattice coupling and bare quark masses are based
             on the work of the CLS collaboration.}
    \label{tab:ensembles}
\end{table}

In order to control the continuum limit we measured the two-point
correlation functions of the vector current using both local and
exactly-conserved discretizations of the current.
Furthermore, in the case of the local-conserved correlation function, there are two discretizations of the $\lambda=-2$ linear combination, 
where in first case the conserved current is constructed on the lattice site and in the second case on the midpoint of the link.
Therefore, we can define the $\lambda=-2$ correlator at half-integer or
integer units of the lattice spacing, through the following replacements
\begin{align}
    G^{ij}(\tau+a/2,\bm k)&\leftarrow\frac{1}{2}\left(G^{ij}(\tau,\bm k) + G^{ij}(\tau+a,\bm k)\right)\\
    G^{00}(\tau,\bm k)&\leftarrow\frac{1}{2}\left(G^{00}(\tau-a/2,\bm k) + G^{00}(\tau+a/2,\bm k)\right)
\end{align}
This gives in total four independent discretizations of the correlation
function associated with the spectral function of eq.~\eqref{eq:rholambda}:
local-local (LL), local-conserved site-centered (LC site), local-conserved
link-centered (LC link) and conserved-conserved (CC).
The correlation functions were measured with all on- and off-axis spatial momenta
up to $k\beta\lessapprox 2\pi$.

\begin{figure}[h]
    \centering
    \includegraphics[scale=0.4]{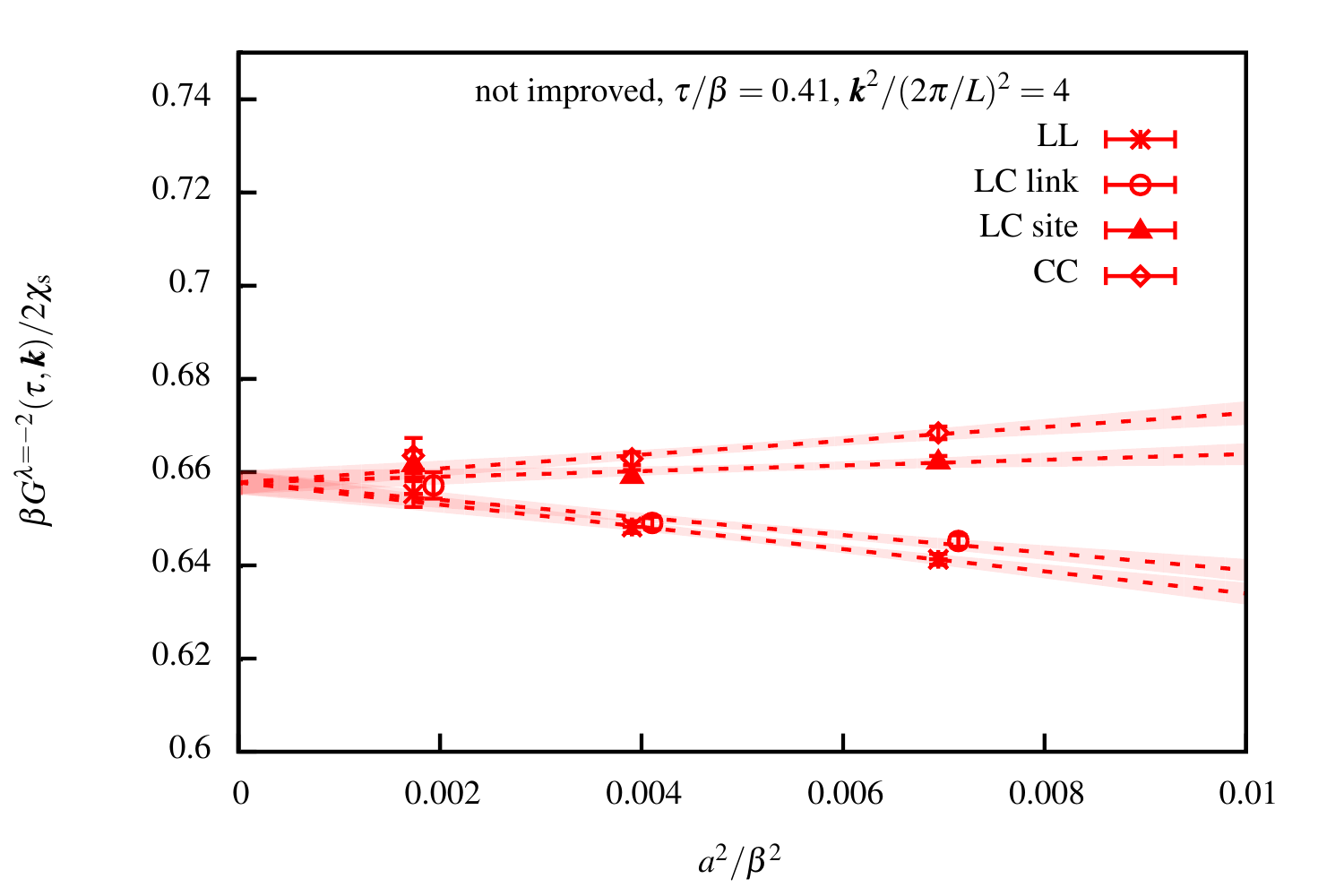}
    \includegraphics[scale=0.4]{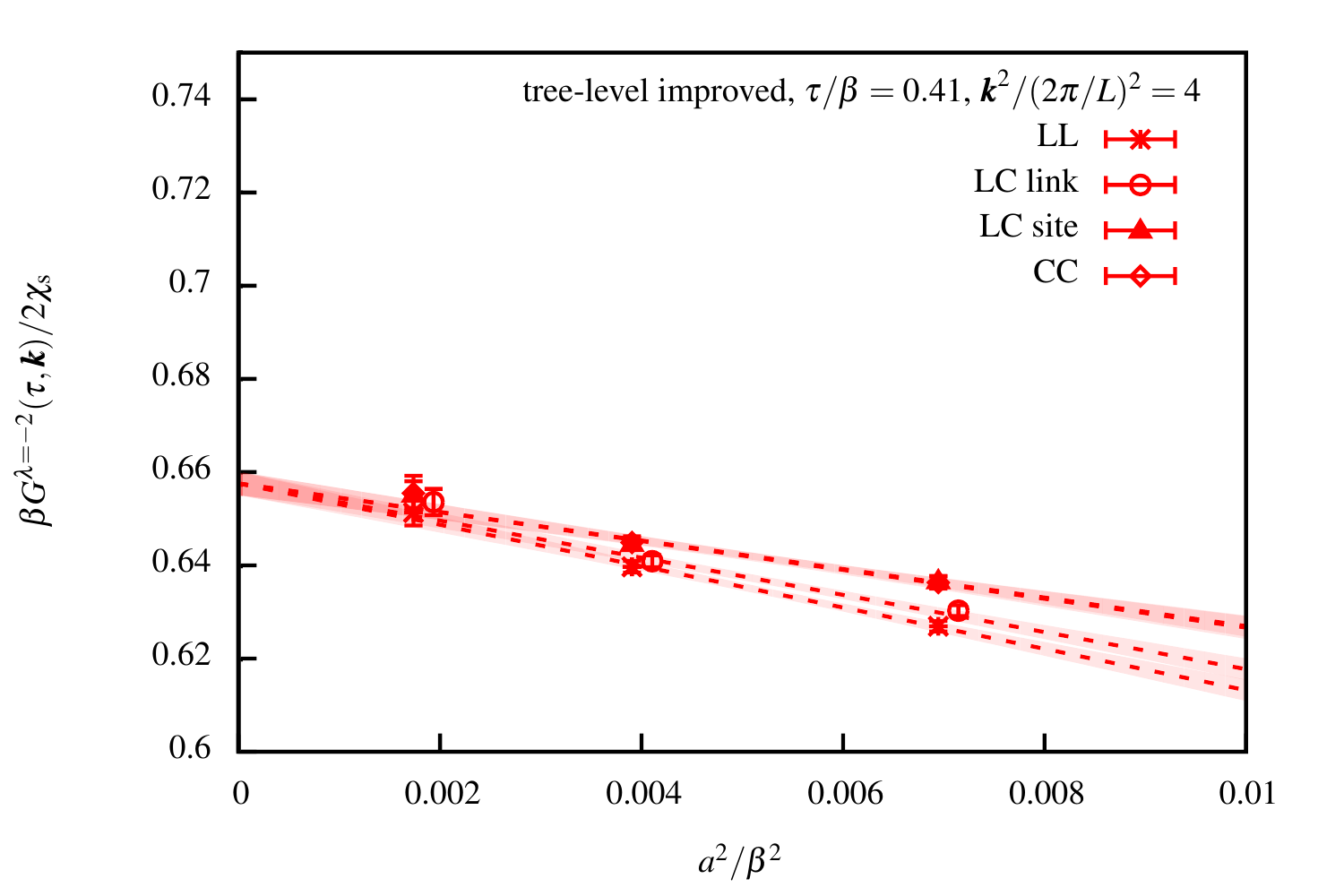}
    \caption{The simultaneous continuum limit of all four discretizations of
             the correlator at a fixed Euclidean distance, momentum and
             temperature.
             The continuum limit without and with tree-level improvement is shown
             in the left and right panels respectively.}
    \label{fig:contlimit}
\end{figure}

The continuum limit was performed by making a piecewise spline interpolation of the
correlation functions and performing a quadratic regression in the lattice
spacing.
This is motivated by the fact that in the chirally-symmetric phase, the matrix
elements of the improvement operators are all suppressed by the quark mass in
units of the temperature, and O($a$)-effects are almost absent.
This behaviour is also clearly observed in the data, as shown in
figure~\ref{fig:contlimit} (left panel) for a given spatial momentum and
Euclidean time separation.
The continuum limit of each of the discretizations agrees, which shows that
discretization effects should be under control, thus making a simultaneous
extrapolation viable.

\begin{figure}[h]
    \centering
    \includegraphics[scale=0.5]{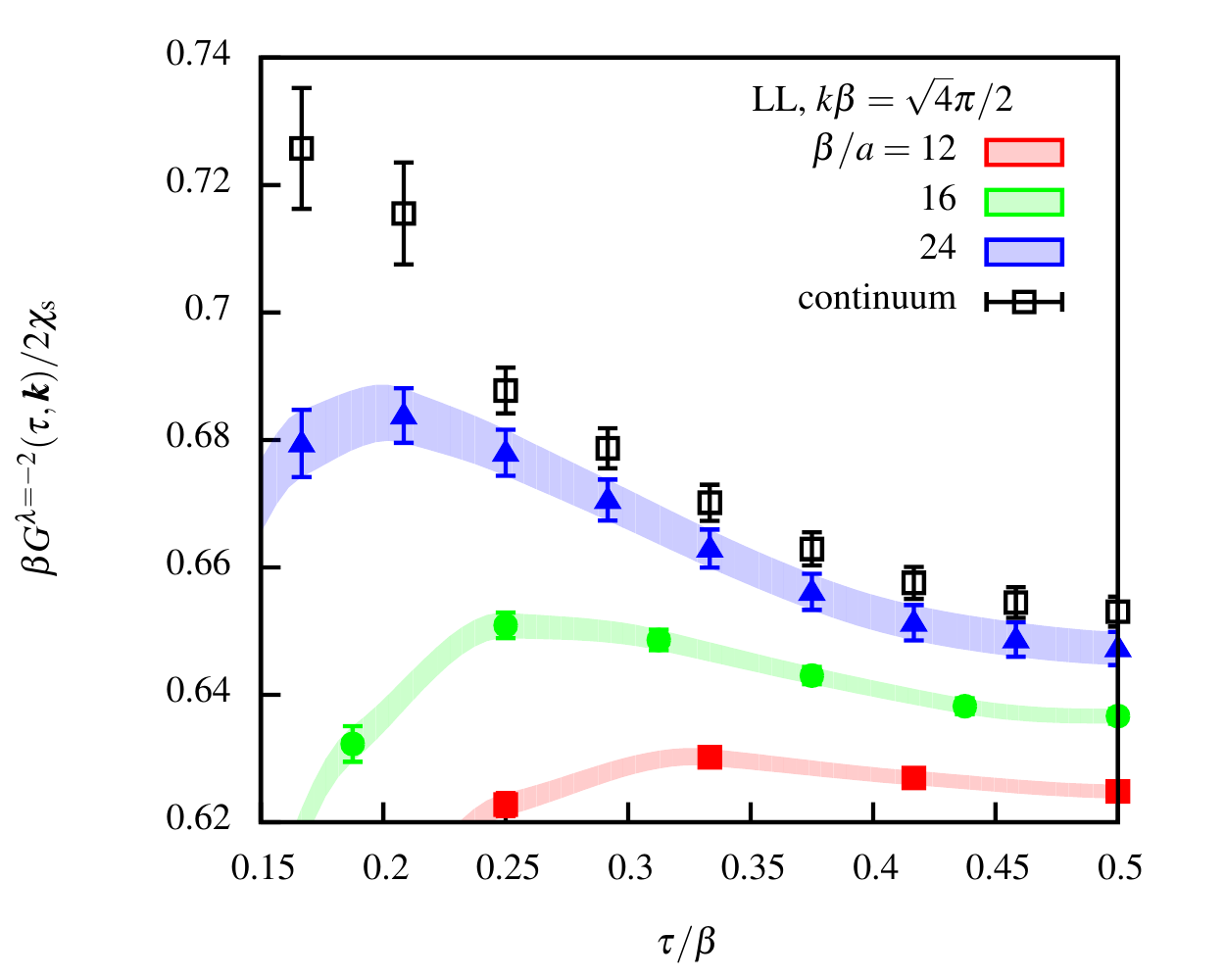}
    \includegraphics[scale=0.5]{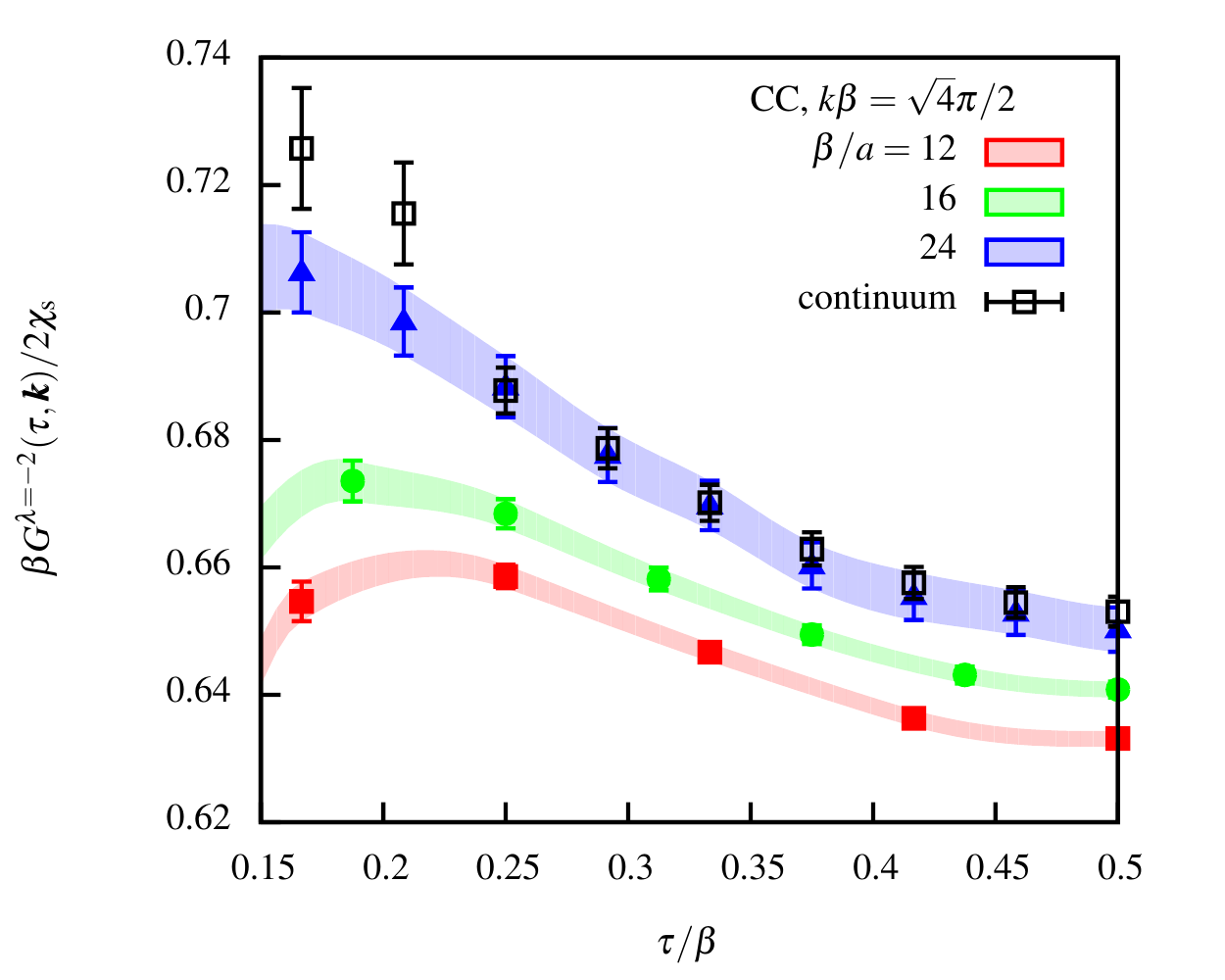}
    \caption{Finite lattice spacing correlators (filled symbols) and
        continuum-extrapolated correlators (open symbols) for the
        LL discretization (left) and the
        CC discretization (right).}
    \label{fig:contcorr}
\end{figure}

Furthermore, we employ an alternative improvement prescription by
multiplicatively removing the tree-level lattice artifacts, via
\begin{align}
        G^{\lambda=-2}(\tau,\bm k) \rightarrow 
        \frac{G^{\lambda=-2}_{\mathrm{cont. t.l.}}(\tau,\bm k)}
             {G^{\lambda=-2}_{\mathrm{lat. t.l}}(\tau,\bm k)}
        G^{\lambda=-2}(\tau,\bm k).
    \end{align}
The effect of this improvement is shown in figure~\ref{fig:contlimit} (right
panel).
Although this provides a substantial correction to the finite lattice spacing
data, reassuringly it does not alter the continuum limit.
The continuum-extrapolated correlator is displayed as the open black symbols in
figure~\ref{fig:contcorr}, which also shows the data away from the continuum with
the closed coloured symbols.
The extrapolation is constrained by the data in both
panels and also the other two discretizations not shown.
Except at short Euclidean distances $\tau\lessapprox\beta/4$, the $\beta/a=24$ correlator
lies close to the continuum-extrapolated values.
The addition of this large and fine lattice is therefore very valuable for the
analysis of cutoff effects.

\section{Analysis of the photon rate}
\label{sec:analysis}

We employ two methods to regularize the inverse problem in the
finite-dimensional analogue of eq.~\eqref{eq:specrep} in order to analyze the continuum-extrapolated correlators 
in terms of the spectral function.
The first is a linear method which gives a model-independent estimator for a smeared spectral function, and the second is to 
fit the parameters of an explicit model to the data.
These are complementary approaches because exact constraints that the
spectral function must satisfy can be built into the model and checked a
posteriori for the linear method.

\subsection{Backus-Gilbert method}
\label{ssec:bgmethod}
The Backus-Gilbert (BG) method~\cite{GJI:GJI169} is defined by constructing a linear map from the space of
functions in the time domain, $G$, to the space of functions on the frequency domain,
$\rho_\mathrm{BG}$,
\begin{align}
    \rho_\mathrm{BG}(\omega,\bm k) &\equiv \sum_n q_n(\bar\omega)G(na,\bm k)\nonumber\\
    &= \int_0^\infty\frac{\mathrm d\omega}{2\pi}\,\rho(\omega,\bm k)
\sum_n        q_n(\bar\omega)K(na,\omega)
\end{align}
by choosing the factor, $\sum_n
q_n(\bar\omega)K(na,\omega)\equiv\hat\delta(\bar\omega,\omega)$, called the
{resolution function}, to be an optimally localized kernel in the frequency domain.
In the case of a constant spectral function the BG estimator coincides with
the spectral function, which explains the advantage of a slowly varying spectral
function.
Other kernels can be constructed with a variety of properties, at the loss of
resolution in the frequency domain.

\begin{figure}[h]
    \centering
    \includegraphics[scale=0.7]{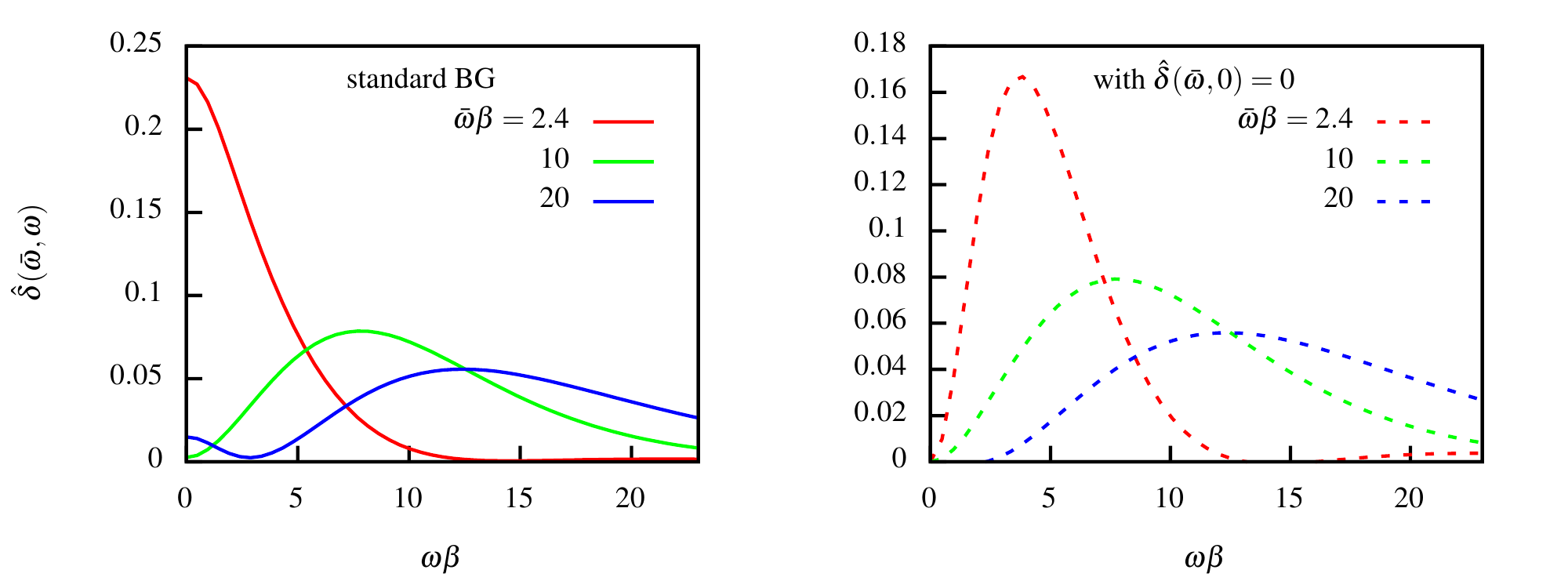}
    \caption{The resolution functions for the standard BG method (left)
             centered at three different frequencies $\bar\omega\beta=2.4,10,20$
             are shown with solid lines.
             The same for the BG method with the extra constraint
             $\hat\delta(\bar\omega,0)=0$.}
    \label{fig:bgres}
\end{figure}

In figure~\ref{fig:bgres} (left), the resolution functions are shown at a selection
of values of the first argument, $\bar\omega$.
Due to the finite number of data, in practice the resolution functions have support on
the whole frequency domain, and for this temperature, are far from
being close to Dirac distributions.
This means that the estimator $\rho_\mathrm{BG}(\bar\omega)$
picks up contributions from the true spectral function at nearby frequencies,
and so is smeared out.
Generally, the resolution decreases at larger values of $\bar\omega$.

In the right-hand panel of figure~\ref{fig:bgres}, an extra linear constraint is introduced,
$\hat\delta(\bar\omega,0)=0$, which means that the estimator obtained
from this method does not contain contributions from the true spectral
function at $\omega=0$.
At small $k$, hydrodynamics predicts a zero mode in the charge-charge correlator
which could give a large contribution to the BG estimator in the
region of the origin.
We use this method to suppress this contribution to the photon rate,
and to check the influence of this diffusion pole on our results.
For larger frequencies, $\bar\omega$, this constraint is irrelevant, but for
the lowest frequency shown, $\bar\omega\beta=2.4$, the support of the
resolution function is pushed away from the origin.

\begin{figure}[h]
    \centering
    \includegraphics[scale=0.551]{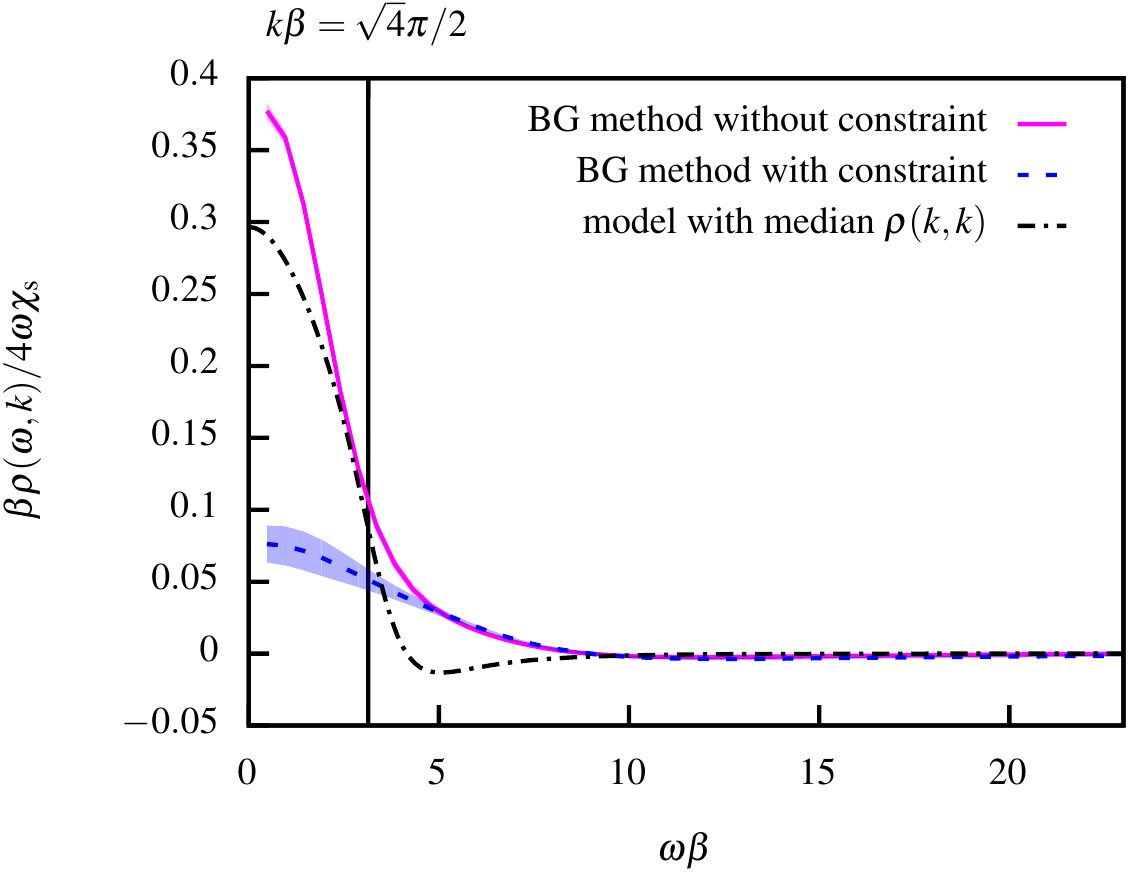}\quad\quad
    \includegraphics[scale=0.551]{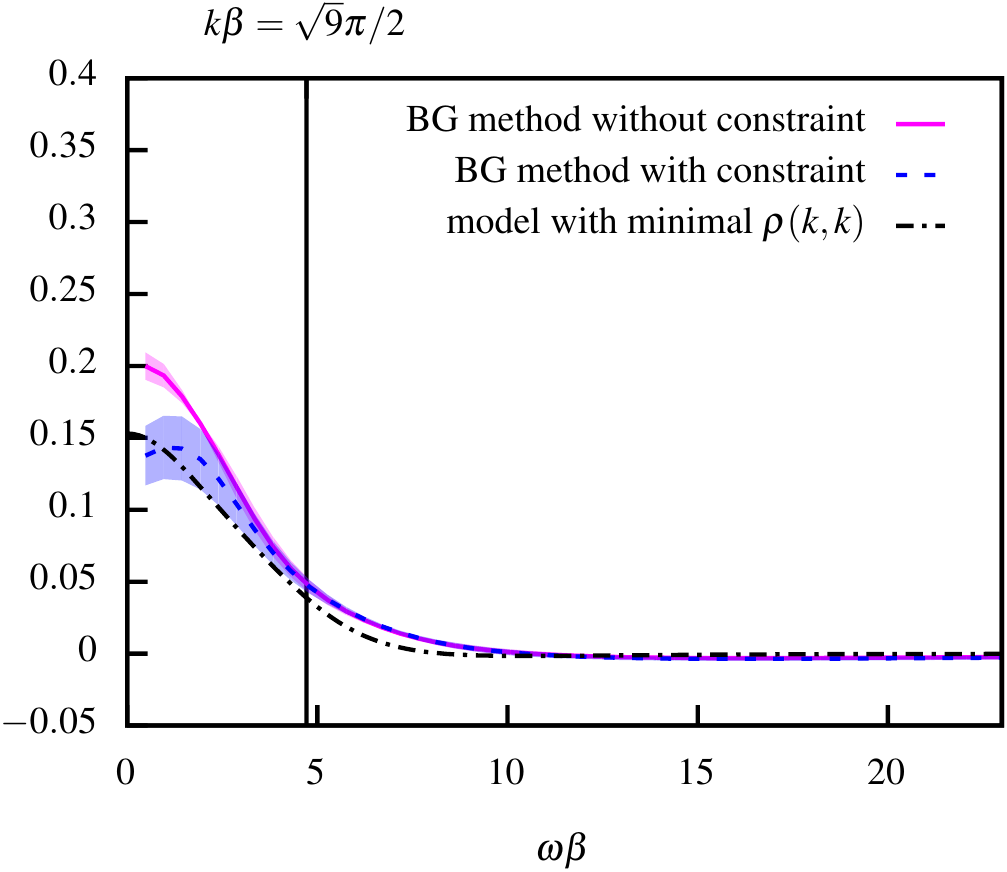}
    \caption{The estimators $\rho_\mathrm{BG}(\omega,\bm k)$ for
             $k\beta=\pi,3\pi/2$ in the left and right panels with the solid pink lines. The solid vertical bars indicate the light-cone frequency in each
             case. The dashed blue curves show the estimator constructed with the extra
             constraint $\hat\delta(\bar\omega,0)=0$. The dashed-dotted black line depicts the model estimate with the median/minimal value of the effective diffusion constant, for a discussion see subsubsection \ref{sssec:compareBGtoFit}.}
    \label{fig:rhobg}
\end{figure}

The BG estimator, $\rho_\mathrm{BG}(\bar\omega)$, is shown for
$T=250\,\textrm{MeV}$ in figure~\ref{fig:rhobg} for $k\beta=\pi$ (left) and
$k\beta=3\pi/2$ (right) panel, normalized to give the \emph{effective}
diffusion constant, $D_\mathrm{eff}/\beta$, on the light-cone as defined in
ref.~\cite{Ghiglieri:2016tvj},
\begin{align}\label{eq:effdiffconst}
    D_\mathrm{eff}(k) =\frac{\rho_\mathrm{BG}(k,k)}{4k\chi_\mathrm{s}},
\end{align}
where $\chi_\mathrm{s}=\int {\rm d^4} x \langle V^\mathrm{em}_0(x) V^\mathrm{em}_0(0)\rangle$, which we know in the continuum as well.
The light-cone frequency is indicated with the vertical solid bar.
The estimator from the BG method with the extra constraint
$\hat\delta(\bar\omega,0)=0$ is shown with the dashed blue curves.
At the lower photon momentum, there is a systematic difference between the two
estimators for the effective diffusion constant.
This might be expected from increasing contributions from the diffusion pole
in the hydrodynamic limit.
On the other hand, for larger photon momenta, this effect is negligible, as
seen in the right-hand panel of the figure.

\begin{figure}[h]
    \centering
    \includegraphics[scale=0.4]{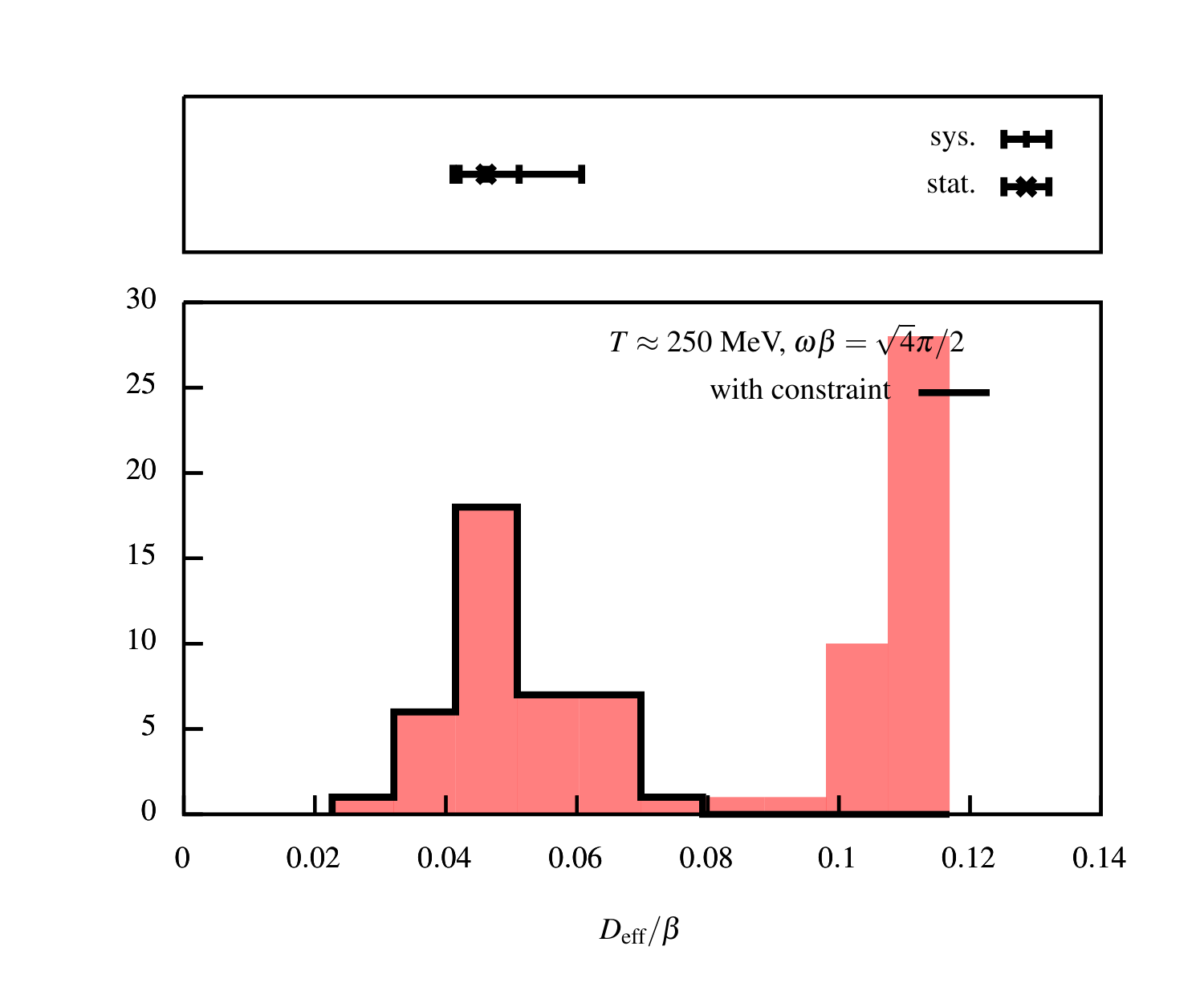}
    \includegraphics[scale=0.4]{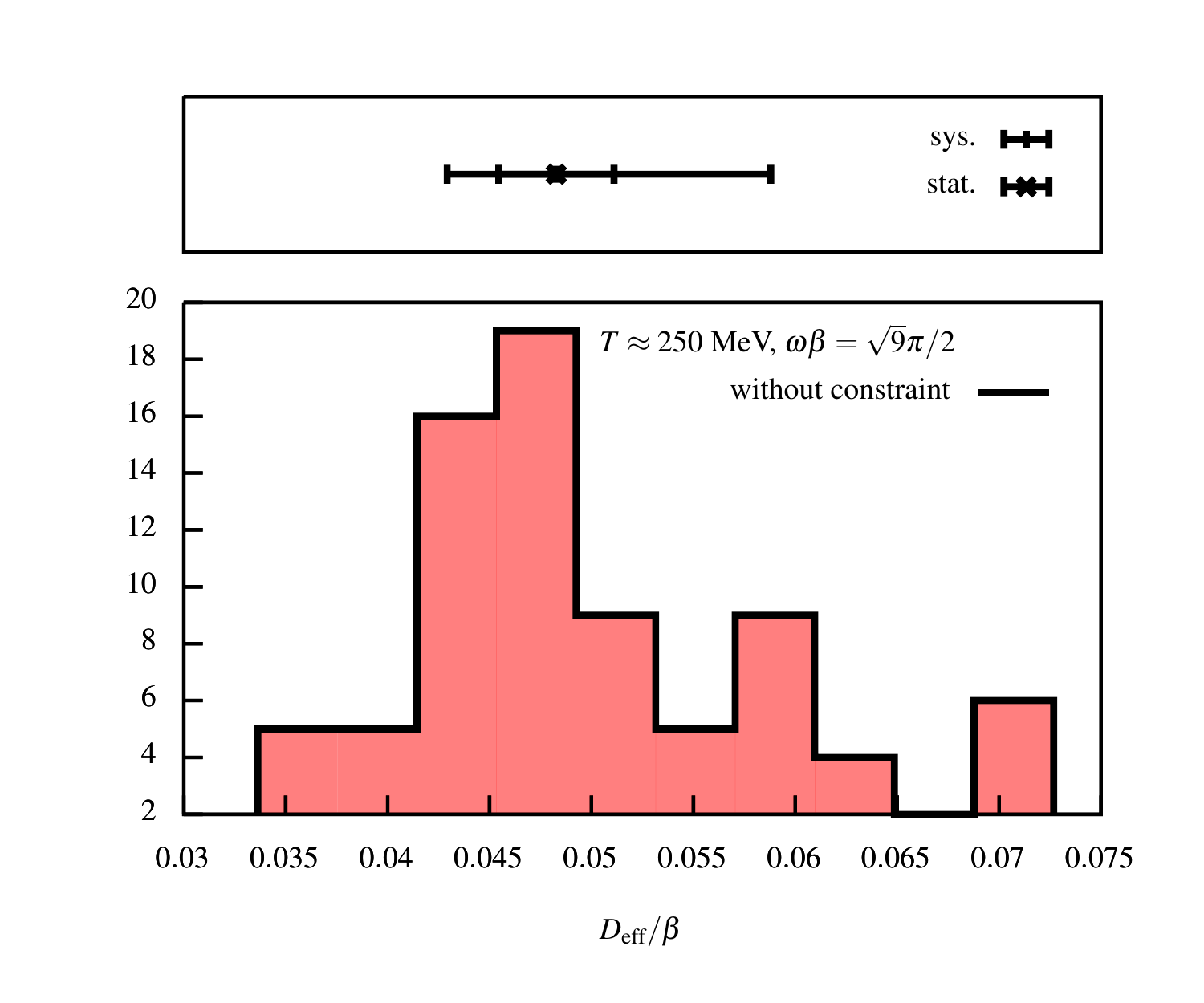}
    \caption{Histograms of the effective diffusion constant for $k\beta=\pi$
        (left) and $k\beta=3\pi/2$ for $T=250\,\mathrm{MeV}$. The resulting
    estimator for the histogram in the black line is shown in the upper
panels, with the inner error bar the statistical error on the median, and the
outer error bar the 68\% intervals, which represents a conservative systematic
error under all the variations of the method.}
    \label{fig:deffhist}
\end{figure}

In order to obtain an estimate for the systematic uncertainty on the effective
diffusion constant from the BG method, we perfomed many variations
of the method which give rise to systematic errors, including variations of
the minimum time separation of the correlator and using the extra constraint.
All the variations are listed in table~\ref{tab:bg_variation}.
In figure~\ref{fig:deffhist}, the distributions are shown for two photon
momenta, $k\beta=\pi,3\pi/2$ in the left and right panels respectively.
For the lower momenta (left), implementing the extra constraint results in a bimodal
distribution, and the two estimators are clearly incomptabile.
In the following the results from this variation are kept separate.
At larger momenta (right), the results become compatible which indicates that
contributions from the diffusion pole are absent in the estimator for the
effective diffusion constant, and therefore combined to obtain the final
estimate of the sytematic uncertainty.
As a final estimator, we take the median of the distribution of the effective
diffusion constant under all variations, and as a systematic uncertainty the 68\%
interval.

\begin{table}
    \centering
    \begin{tabular}{lr}
        \toprule
        variation   &   values \\
        \midrule
        $\tau_\mathrm{min}/\beta$ &  $ \{0.1,\ldots,0.25\}$ \\
        with $\hat\delta(\bar\omega,0)=0$ & \{yes, no\}\\
        BG regularization parameter $\alpha$ &   $\{10^{-2},\cdots,10^{-4}\}$ \\
        tree-level improved &\{yes, no\}\\
        \bottomrule
    \end{tabular}
    \caption{Variations of the BG method used to estimate a systematic
    uncertainty.}
    \label{tab:bg_variation}
\end{table}

\subsection{Model Fit}
\label{ssec:fitansatz}

In the previous subsection \ref{ssec:bgmethod}, a model-independent approach is discussed, the Backus-Gilbert method. To complement this approach we study a physics-inspired ansatz for the $\tanh$-regulated spectral function in this subsection. Thus, we are able to perform a maximum likelihood estimation by a nonlinear fit. Before discussing the ansatz we first prove some analytic properties of the spectral function which will motivate and strongly constrain the choice of simple models we can use for a maximum likelihood estimator.

\subsubsection{Sum rule}
\label{sssec:sumrule}
In vacuum, Lorentz invariance and transversity of $G_{\mu\nu}(\tau,\bm k)$ result in
\begin{eqnarray}
 \rho_{\lambda=-2}(\omega, \bm k)=\left(\delta^{ij}-\frac{3 k^i k^j}{k^2}\right)\rho_{ij}(\omega,\bm k)+2\rho_{00}(\omega,\bm k)=0
\end{eqnarray}
and at $T>0$, this combination is still UV finite as no new divergences appear. This admits an operator product expansion (OPE) for the spectral function: by power counting, $\rho_{\lambda=-2}(\omega,\bm k) \sim \frac{\left\langle\mathcal{O}_4\right\rangle}{\omega^2}+\cdots$ as $\omega\rightarrow\infty$, where $\mathcal O_4$ is a dimension-four operator.
Additionally, $\rho_{\lambda=-2}(\omega,\bm k= \bm 0)=0$ at $\omega>0$ due to charge conservation and therefore we expect from the OPE
\begin{eqnarray}\label{eq:OPErholambda}
 \rho_{\lambda=-2}(\omega, \bm k)\propto \frac{k^2\left\langle\mathcal O_4\right\rangle}{\omega^4}, \quad\quad \omega \gg \pi T,k.
\end{eqnarray}

In momentum space, we can expand the spectral representation of the Euclidean correlator about $\omega_n=\infty$,
\begin{eqnarray}\label{eq:momspacerholambda}
 \tilde G(\omega_n,\bm k)&=&\int_0^\infty\frac{\rm d\omega}{\pi}\,\omega\,\frac{\rho_{\lambda=-2}(\omega,\bm k)}{\omega^2+\omega_n^2}\nonumber\\
 &\stackrel{\omega_n\rightarrow\infty}{\longrightarrow}&\frac{1}{\pi\,\omega_n^2}\int_0^\infty {\rm d} \omega\,\omega\,\rho_{\lambda=-2}(\omega,\bm k)+\mathcal{O}(\omega_n^{-4}).
\end{eqnarray}
The OPE, however, tells us that the first non-vanishing contribution is of $\mathcal{O}(\omega^{-4})$, see eq. \eqref{eq:OPErholambda}. Thus, the first coefficient in eq. \eqref{eq:momspacerholambda} must vanish exactly, and we deduce the superconvergent sum rule
\begin{eqnarray}\label{eq:sumrule}
 \int_0^\infty{\rm d} \omega\,\omega\,\rho_{\lambda=-2}(\omega,\bm k)=0.
\end{eqnarray}

\subsubsection{Pad\'e ansatz}
\label{sssec:padeansatz}
The $\tanh$-regulated spectral function can be described by a combination of two Pad\'e approximants as
\begin{eqnarray}\label{eq:Padeansatz}
   \frac{\rho(\omega,k)}{\tanh(\omega\beta/2)}=\underbrace{\frac{A}{\left[\omega^2+a^2\right]}}_{\text{part I}} \cdot \underbrace{\frac{(1+B\omega^2)}{\left[(\omega+\omega_0)^2+b^2\right]\left[(\omega-\omega_0)^2+b^2\right]}}_{\text{part II}}
\end{eqnarray}
with two linear parameters $A$ and $B$ as well as three nonlinear parameters $a$, $\omega_0$, and $b$.
Part I is inspired by the diffusion pole as it arises in the hydrodynamics prediction for the infrared limit; when one identifies $a\leftrightarrow Dk^2$ for small $k$, part I resembles the known expression \cite{CaronHuot:2006te}
\begin{eqnarray}
 \frac{\rho(\omega,k)}{\omega}\approx\frac{4\chi_s Dk^2}{\omega^2+(Dk^2)^2}, \quad\quad \omega,k\ll D^{-1}.
\end{eqnarray}
Part II models the pole structure of the AdS/CFT current correlator, see ref. \cite{Kovtun:2005ev} for details.

\begin{figure}[t]
    \centering
    \includegraphics[scale=0.3]{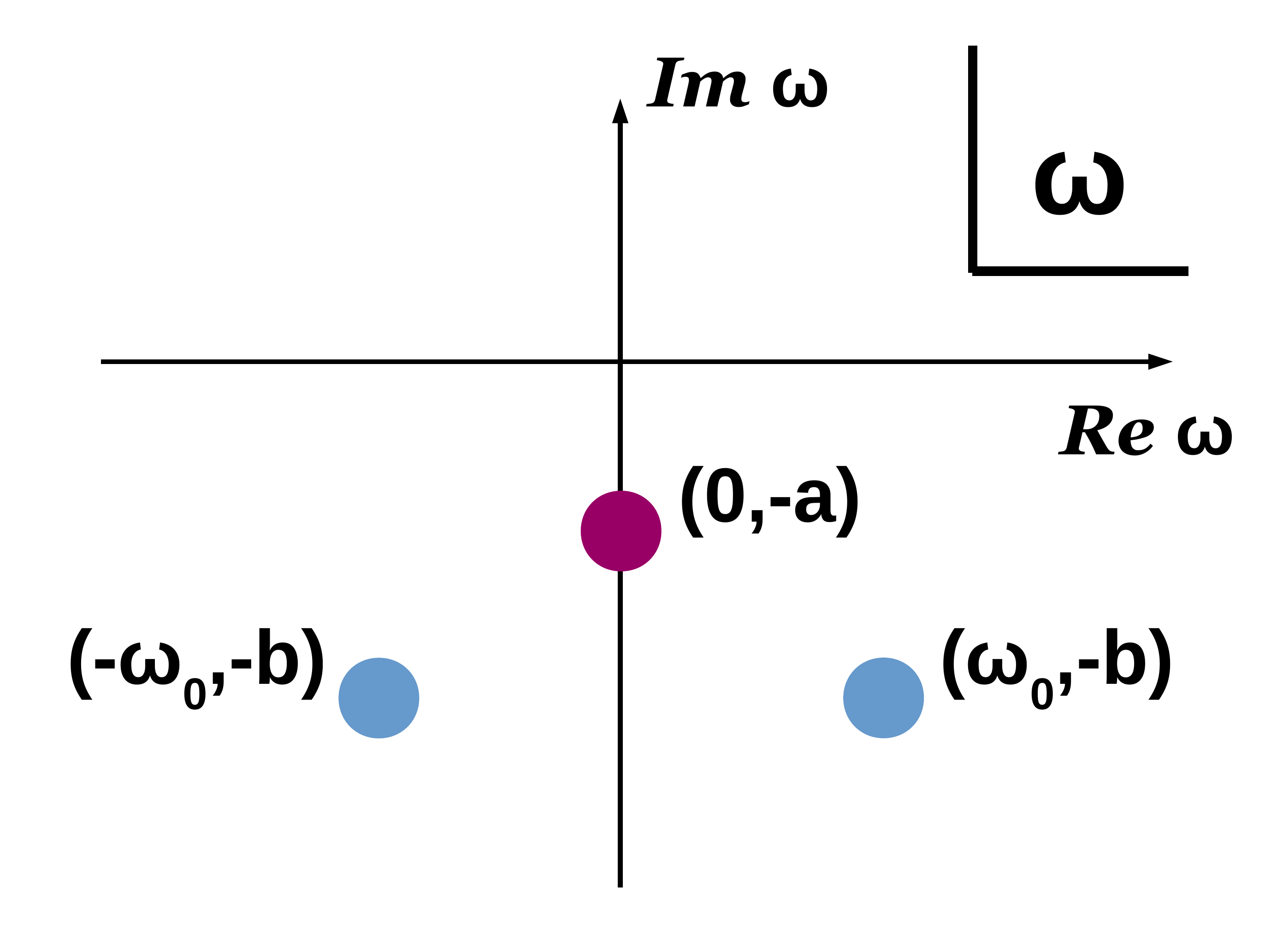}
    \caption{Complex $\omega$-plane with the poles of the retarded correlator. There is the diffusion pole on the imaginary axis (red) and a pair of generic poles in the lower half-plane (blue) which model the pole structure of the AdS/CFT current correlator.}
    \label{fig:complexplane}
\end{figure}

In order to satisfy the superconvergent sum rule \eqref{eq:sumrule}, a second linearly independent parameter $B$ has to be introduced in part II of eq. \eqref{eq:Padeansatz}. By imposing eq. \eqref{eq:sumrule}, the second linear parameter $B$ becomes a function of $(a,\omega_0,b)$. The two poles at $(\pm \omega_0,-b)$ in the lower half-plane do not only model the quasinormal modes of the retarded correlator, discussed in ref. \cite{Kovtun:2005ev}, they also match the $1/\omega^4$ behaviour at large $\omega$ as dictated by the operator product expansion \eqref{eq:OPErholambda}, see also figure \ref{fig:complexplane}.

\subsubsection{Uncorrelated fit}
\label{sssec:uncorrfit}
\begin{figure}[t]
    \centering
    \includegraphics[scale=0.4]{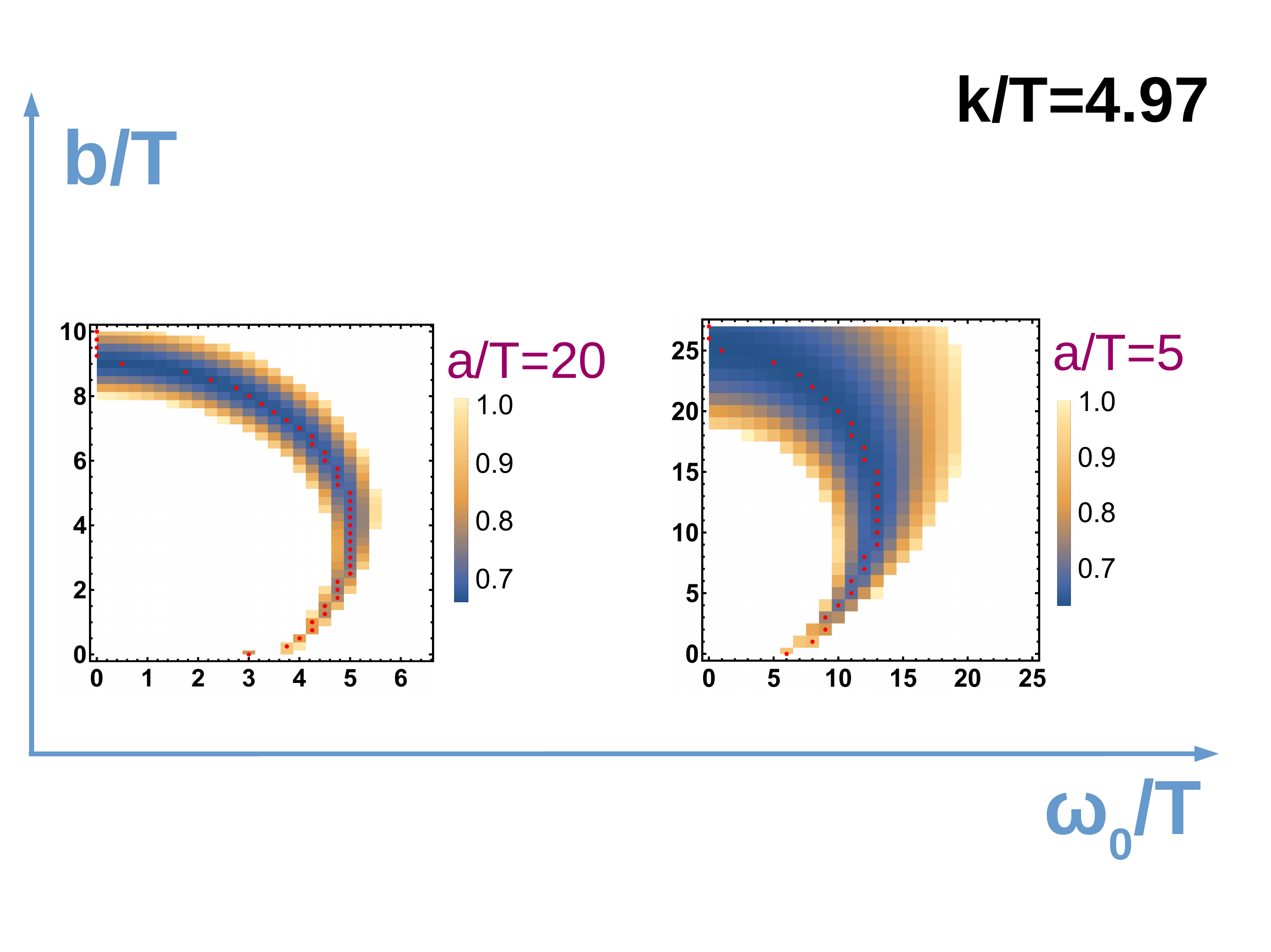}
    \caption{Uncorrelated $\chi^2$-landscape in $(\omega_0,b)$-plane at photon momentum $k/T\approx 4.97$ for fixed values of $a$. The valley of acceptable $\chi^2<1$ is shown whereas the area with $\chi^2>1$ is marked in white. The red dots lie at nearly degenerate minima at the bottom of the $\chi^2$-valley. When the diffusion pole is fixed to be far from the origin, $a/T=20$ (left panel), the generic pole at $(\omega_0,b)$ can move close to the real axis. In case the diffusion pole dominates the shape of the spectral function, $a/T=5$ (right panel), the second pole is pushed into the complex plane.}
    \label{fig:chisqlandscape}
\end{figure}

In the fit ansatz \eqref{eq:Padeansatz}, we are left with four independent fit parameters after determining $B$ for given $(a,\omega_0,b)$ via the sum rule \eqref{eq:sumrule}. Then there are three degrees of freedom when we use seven data points between $\tau_{\mathrm{min}}/\beta=0.25$ and $\tau_{\mathrm{max}}/\beta=0.5$ on the continuum-extrapolated correlators. There is a huge subspace in the parameter space $(A,a,\omega_0,b)$ for which the uncorrelated $\chi^2$ is smaller than one. Figure \ref{fig:chisqlandscape} depicts the $\chi^2$-valley in the $(\omega_0,b)$-plane at photon momentum $k/T\approx 4.97$ where $\chi^2<1$ at two fixed values of $a$.

Because there is no local minimum in the uncorrelated $\chi^2$-landscape, this results in a plethora of acceptable solutions. In other words, there are many shapes of the spectral function, none of which can be strongly ruled out by our data as they all satisfy $\chi^2(A,a,\omega_0,b)<1$. So rather than minimizing the uncorrelated $\chi^2$, we try and find bounds to the effective diffusion constant by taking the \emph{min} and \emph{max} values of  all photon rates with $\chi^2(A,a,\omega_0,b)<1$.

\begin{figure}[t]
    \centering
    \includegraphics[scale=0.7]{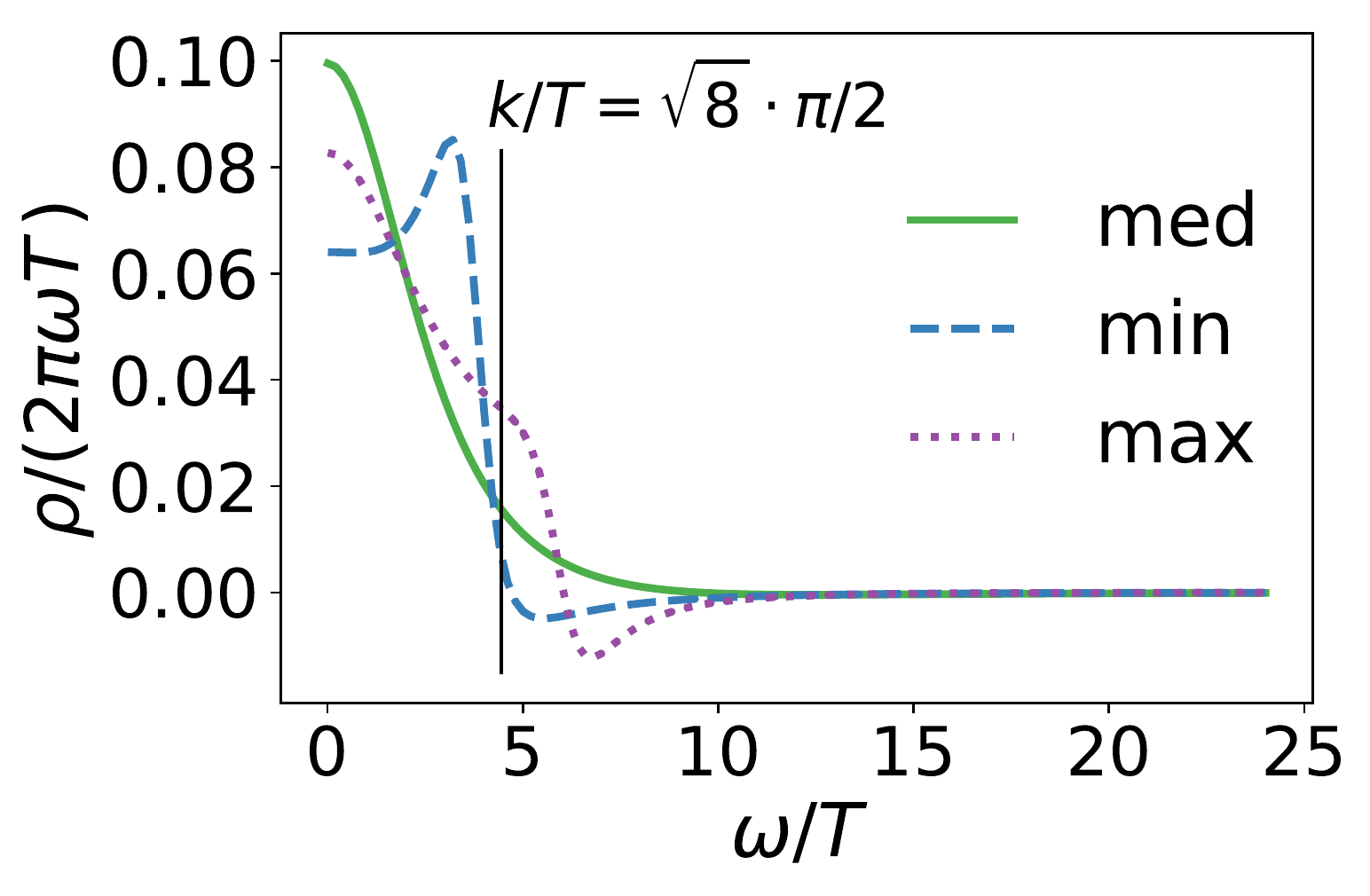}
    \caption{The three different spectral functions are solutions with $\chi^2<1$ and result in the median (solid green), \emph{min} (dashed blue) and \emph{max} (dotted purple) value of the photon rate distribution. The vertical black line marks the light cone at photon momentum $k/T=\sqrt{8}\cdot\pi/2\approx 4.44$. There is a pronounced peak left of the light cone for the dashed blue curve, a slightly less pronounced peak right of the light cone for the dotted purple line and no additional peak apart from the diffusion pole for the solid green curve. The spread between the minimal and maximal photon rate results in the spread of the diffusion constant.}
    \label{fig:specfunc}
\end{figure}

We exclude all solutions that result in a negative photon rate from this procedure.
This is due to the known positivity of the spectral function below the light cone,
\begin{eqnarray}
 \rho(\omega)\geq 0, \quad\quad \omega\leq k.
\end{eqnarray}
When one allows the second pole at $(\omega_0,b)$ to get too close to the real axis, the result is a very pronounced peak in the spectral function. This can happen when the nonlinear parameter $a$ becomes large and the diffusion peak does not dominate the analyticity of the retarded correlator. For illustration see figure
\ref{fig:chisqlandscape}: at $a/T=20$ (left panel) the diffusion pole is far
from the origin and the second pole at $(\omega_0,b)$ can approach the real axis. In figure \ref{fig:specfunc}, the result is seen as a sharp peak below the light cone (dashed blue curve). For $a/T=5$ (right panel of figure \ref{fig:chisqlandscape}), however, the second pole at $(\omega_0,b)$ is pushed into the complex plane and does not dominate the shape of the spectral function. 

Such an additional peak from a pole close to the real axis corresponds to a
very long-lived excitation in the medium which is unphysical if it survives longer than the largest possible relaxation times in the system. So we constrain the imaginary parts of both poles, i.e. the fit parameters $a$ and $b$, to fulfil the exclusion criterion \cite{CaronHuot:2006te,Arnold:2003zc}
\begin{eqnarray}\label{eq:exclusioncrit}
 \min(a,b)>\min(D_{\mathrm{AdS/CFT}}\cdot k^2,D_{\mathrm{PT}}^{-1}),
\end{eqnarray}
where $D_{\mathrm{AdS/CFT}}\cdot k^2$ describes the diffusion of an electric
charge with $D_{\mathrm{AdS/CFT}}=1/(2\pi T)$ and $D_{\mathrm{PT}}^{-1}$
accounts for the damping of a static current with
$D_{\mathrm{PT}}^{-1}=\mathcal{O}(\alpha_s^2)\cdot T$, $\alpha_s=0.25$. We
claim that the exclusion criterion \eqref{eq:exclusioncrit} amounts to a conservative constraint based on physics considerations.

\subsubsection{Comparison of BG method and Pad\'e fit}
\label{sssec:compareBGtoFit}
Figure \ref{fig:rhobg} compares the BG estimators with and without the
constraint on the resolution function with the model ansatz
\eqref{eq:Padeansatz} for two different photon momenta $k\beta=\pi$ and
$3\pi/2$ in the left and right panel, respectively. For $k\beta=\pi$ the
biggest discrepancy between the BG method and Pad\'e ansatz occurs below the light cone for the BG estimator with the constraint. On the light cone and above, the discrepancy is small regardless of whether the resolution function is constrained or not. For $k\beta=3\pi/2$ there is almost no systematic difference between the BG estimators with or without a constraint and the fit ansatz. An important point to note is that on the light cone there is very good agreement of the BG estimators and the median of the model distribution in all cases.

\section{Results}
\label{sec:results}
Having discussed and compared the two approaches in the previous section
\ref{sec:analysis}, we can now plot the effective diffusion constant,
eq.~\eqref{eq:effdiffconst}, as a function of photon momentum $k$.

\begin{figure}[t]
    \centering
    \includegraphics[width=0.8\textwidth]{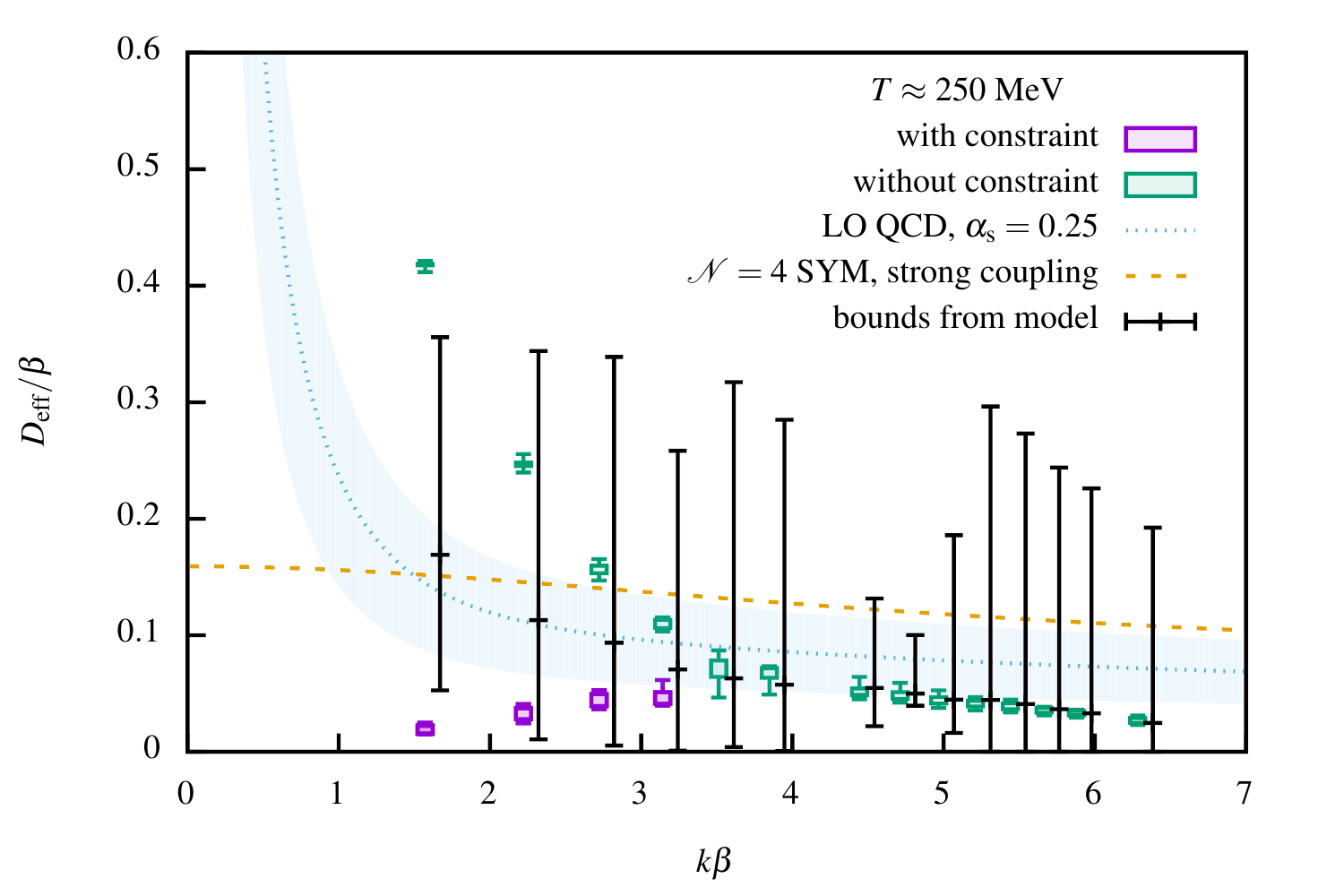}
    \caption{Estimate of the effective diffusion constant at $T\approx 250\mathrm{MeV}$. The results from the BG method are plotted as purple and green dots. The bounds from the model and the median of the distribution are displayed as black bars. Additionally, the strong-coupling result from $\mathcal{N}=4$ SYM and a weak-coupling result from leading-order (LO) perturbative QCD with $\alpha_{\mathrm{s}}=0.25$ are shown.}
    \label{fig:deff_T250MeV}
\end{figure}

Figure \ref{fig:deff_T250MeV} shows the results both from the BG method and the spread from the distribution of solutions to the uncorrelated fit at $T=250\,\mathrm{MeV}$.
As discussed in subsection \ref{ssec:bgmethod}, the two BG estimators resulting from implementing or not implementing the constraint are not compatible with each other at lower momenta and the two results are plotted separately (green vs. purple dots). At higher momenta they become compatible and only one distribution is considered to estimate a systematic and statistical error. The black bars indicate the bounds we find from quoting the minimal and maximal value of the distribution of the effective diffusion constant that have $\chi^2<1$ and the median of this distribution. Furthermore, the strong-coupling result from $\mathcal{N}=4$ supersymmetric Yang-Mills theory (SYM) and a weak-coupling result from leading-order (LO) perturbative QCD with $\alpha_{\mathrm{s}}=0.25$ are plotted as comparison.

For photon momenta $k\beta\gtrsim 3.5$ the median of the uncorrelated fit distribution and the BG estimator are remarkably close and agree within the systematic and statistical error of the BG estimator. At the lower momenta the median of the distribution lies between the two BG estimators with and without the constraint. For the whole momentum range, the bounds of the effective diffusion constant from examining the minimum and maximum values of the maximum likelihood estimator consistent with the data, cover a big interval and it is not possible to discriminate between the weak-coupling and the strong-coupling scenarios. At lower momenta the spread from the model is compatible with the separation of the two BG estimators.

\begin{figure}[t]
    \centering
    \includegraphics[width=0.8\textwidth]{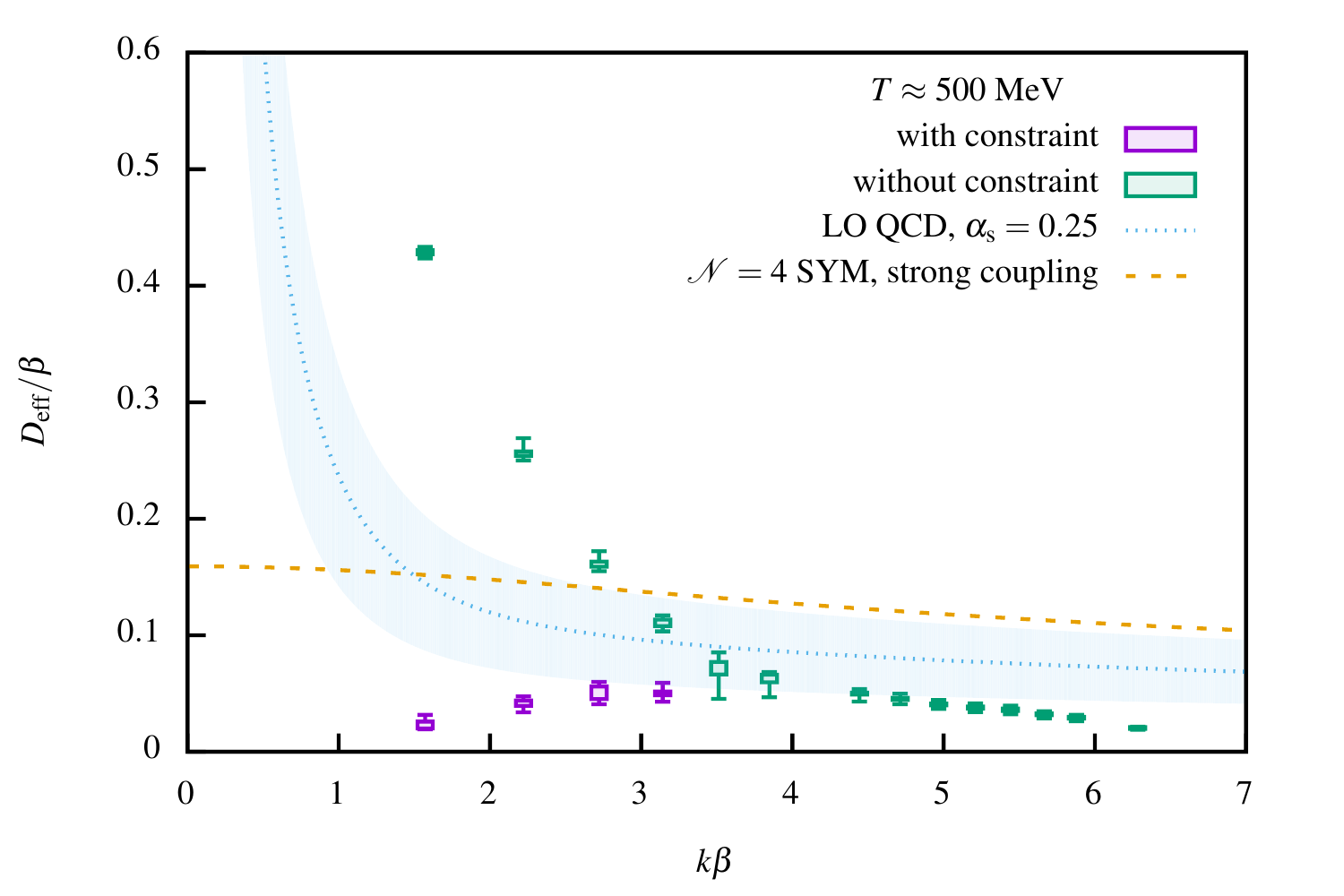}
    \caption{Estimate of the effective diffusion constant at $T\approx 500\,\mathrm{MeV}$ for a single lattice spacing. The results from the BG method are plotted as purple and green dots. Additionally, the strong-coupling result from $\mathcal{N}=4$ SYM and a weak-coupling result from leading-order (LO) perturbative QCD with $\alpha_{\mathrm{s}}=0.25$ are shown. The model has not been fitted to the data at this temperature.}
    \label{fig:deff_T500MeV}
\end{figure}

Figure \ref{fig:deff_T500MeV} shows the estimate of the effective diffusion constant deep in the deconfined phase at $T\approx 500\,\mathrm{MeV}$. For this temperature, we only have one ensemble at a single lattice spacing so there is no continuum extrapolation available, and the Pad\'e ansatz has not yet been analyzed for this data. By comparison with figure \ref{fig:deff_T250MeV}, however, we do not observe any strong temperature dependence of this observable.

\section{Summary and Outlook}
\label{sec:sumout}

We presented an estimate of the photon rate from dynamical QCD based on continuum-extrapolated correlators. In order to be more sensitive to the physics on the light cone, we propose an alternative linear combination of the vector-vector correlator that eliminates the UV contamination.

We avail of two qualitatively different approaches to estimate the photon rate: the Backus-Gilbert method is a linear mapping that reconstructs a smeared estimator of the true spectral function; and a Pad\'e fit ansatz serves as a model inspired by relativistic hydrodynamics, AdS/CFT and plausibility constraints. We can exploit the UV behaviour of the spectral function expected from an OPE and derive a superconvergent sum rule that is implemented into the model.

As the uncorrelated $\chi^2$-landscape is rather degenerate, we quote the median of the distribution of acceptable solutions with $\chi^2<1$ and the \emph{min} and \emph{max} values of this distribution. The median coincides with the BG estimator at larger momenta and is compatible with the separation of the two BG estimators at lower momenta. In order to gain more control in the small $k$ region, it can be useful to define another effective diffusion coefficient $\overline D_{\mathrm{eff}}(\xi,k)=\frac{\xi\rho_{\lambda=-2}(\xi k,k)}{4\chi_{\mathrm{s}} k} \rightarrow D$ which tends to the true diffusion constant $D$ for $k\rightarrow 0$ at fixed $\xi\in\left[0,1\right)$.

A very similar strategy can be applied to energy-momentum tensor correlators. Let $\rho^{\mu\nu,\lambda\sigma}(\omega,\bm k)$ be the spectral function associated with the $\langle T^{\mu\nu}T^{\lambda\sigma}\rangle$ correlator. With the goal in mind to compute the shear viscosity $\eta$, consider, for $\bm k=(0,0,k)$, the linear combination $[\rho^{12,12}-\rho^{13,13}+\rho^{10,10}]$. It vanishes in the vacuum, is positive for ${\omega}^2-k^2<0$ and falls off as $\bm k^2/\omega^2$ at asymptotic frequencies. It is therefore a promising starting point for a calculation of the shear viscosity $\eta$.

Apart from increasing statistics it might be beneficial to further examine our
systematics including the continuum limit. One should also take into account
the correlations between data points as the exclusionary power of the correlated  $\chi^2$ is stronger.
Further cross-checks on the BG estimator can be investigated, such as verifying
how well the sum rule is satisfied.
Finally, extending the study to a higher temperature will illuminate the temperature dependence of the photon rate.

\section*{Acknowledgements}
We acknowledge the use of computing time on the JUGENE and JUQUEEN computers
of the Gauss Centre for Supercomputing located at Forschungszentrum J\"ulich, Germany under grant HMZ21. Part of the calculations were performed on the cluster “Clover” of the Helmholtz-Institute Mainz and on the cluster Mogon II at JGU Mainz. This work was supported by the DFG Grant No. ME 3622/2-2 \emph{QCD at non-zero temperature with Wilson fermions on fine lattices}.

\clearpage
\bibliography{Lattice2017_353_STEINBERG}

\end{document}